\documentclass[aps,pra,
               groupedaddress,
               showpacs,showkeys,
               amsmath,amssymb,amsfonts,
               floatfix,
               a4paper,11pt,onecolumn,oneside,
               final]{revtex4}

\usepackage{hyperref}                            
\usepackage{graphicx}
\usepackage{dcolumn}                             
\usepackage{bm}                                  
\usepackage[verbose,tmargin=3cm,bmargin=3cm,lmargin=3cm,rmargin=3cm]{geometry}

\usepackage[switch,running]{lineno}              

\newcommand{\dif}{\mathrm{d}}                    
\newcommand{\me}{\mathrm{e}}                     
\newcommand{\mpi}{\mathrm{\pi}}                  

\DeclareMathOperator{\erfc}{erfc}                
\DeclareMathOperator{\Real}{Re}                  
\DeclareMathOperator{\Imag}{Im}                  

\begin{document}


\title{\Large
  Photonic superdiffusive motion in resonance line radiation
  trapping - partial frequency redistribution effects}

\author{A.R.~Alves-Pereira}
\author{E.J.~Nunes-Pereira}
\email{epereira@fisica.uminho.pt}
\affiliation{Universidade do Minho, Escola de Ci\^{e}ncias,
  Centro de F\'{i}sica, 4710-057 Braga, Portugal}

\author{J.M.G.~Martinho}
\author{M.N.~Berberan-Santos}
\affiliation{Centro de Qu\'{i}mica-F\'{i}sica Molecular, Instituto
  Superior T\'{e}cnico, 1049-001 Lisboa, Portugal}

\date{\today}

\begin{abstract}

The relation between the jump length probability distribution function and the spectral line profile in resonance atomic
radiation trapping is considered for Partial Frequency Redistribution~(PFR) between absorbed and reemitted radiation. The
single line Opacity Distribution Function~[M.N.~Berberan-Santos~\textit{et.al.} J.~Chem.~Phys. \textbf{125}, 174308, 2006]
is generalized for PFR and used to discuss several possible redistribution mechanisms~(pure Doppler broadening, combined
natural and Doppler broadening and combined Doppler, natural and collisional broadening). It is shown that there are two
coexisting scales with a different behavior: the small scale is controlled by the intricate~PFR details while the large
scale is essentially given by the atom rest frame redistribution asymptotic. The pure Doppler and combined natural, Doppler
and collisional broadening are characterized by both small and large scale superdiffusive L\'{e}vy flight behaviors while
the combined natural and Doppler case has an anomalous small scale behavior but a diffusive large scale asymptotic. The
common practice of assuming complete redistribution in core radiation and frequency coherence in the wings of the spectral
distribution is incompatible with the breakdown of superdiffusion in combined natural and Doppler broadening conditions.

\end{abstract}

\maketitle

\section{INTRODUCTION}

The study of transport phenomena~\cite{Chapman1970} and in particular of diffusion processes~\cite{Mazo2002andHughes1995}
has developed considerably in the course of the last decades. One of the foundations of this area is the well-known
diffusion equation, consequence of the central limit theorem~\cite{Gnedenko1954}, that describes the random motion of a
particle in an isotropic homogeneous three-dimensional space. The Probability Density Function~(PDF) of the diffusing
species being at a certain position~$r$ at time~$t$, after a sufficiently high number of jumps and for an initial location
at~$r = 0$, is given by the propagator~\cite{vanKampen1992,Metzler2000and2004}

\begin{equation}
  \label{eq-DiffusionEq}
  W\left( r,t\right) \propto \frac{1}{\left\langle r^{2}\left( t\right) \right\rangle ^{3/2}}
  \, \exp \left( -\frac{3 r^{2}/2}{\left\langle r^{2}\left( t\right) \right\rangle}\right)
  \mbox{,}
\end{equation}

with the root mean square displacement,

\begin{equation}
  \label{eq-DiffusionCoeff}
  \left\langle r^{2} \left( t \right) \right\rangle = 6Dt
  \mbox{,}
\end{equation}

where $D$ is the diffusion coefficient of the diffusing species.

By the end of the last century it was recognized that the simple behavior in Eqs.~(\ref{eq-DiffusionEq})
and~(\ref{eq-DiffusionCoeff}) was not completely general. The two most widely known deviations from the standard diffusive
behavior are L\'{e}vy flights~\cite{Mandelbrot1983,Shlesinger1995}, coined and popularized by Benoit Maldelbrot, and
Richardson's law of turbulent diffusion~\cite{Shlesinger1987}. Recently, anomalous diffusion equations were also found to
apply to in the econometric modeling and/or prediction~\cite{Bouchaud2000}, in laser cooling of atoms~\cite{Bardou2002}, in
the foraging behavior of some animals~\cite{Viswanathan1996}, and to the the scaling laws ruling human travel
activities~(critically connected to the geographical spreading of infectious diseases)~\cite{Brockmann2006}.

Situations deviating from the classical diffusion case can heuristically be described by a modified power law giving the
scaling of the mean squared displacement as

\begin{equation}
  \label{eq-DiffusionCoeffScaling}
  \left\langle r^{2} \left( t \right) \right\rangle \propto t^{\gamma }
  \mbox{.}
\end{equation}

Anomalous diffusion results in the departure of the $\gamma$ factor from unity in the last equation, owing to the non
applicability of the classical central limit theorem, as a result of the presence of broad distributions or long-range
correlations and non-local effects~\cite{Metzler2000and2004,Bouchaud1990}. Broad spatial jump or waiting time distributions
lead to a non-Gaussian and possibly a non-Markovian time and spatial evolution of the
system~\cite{vanKampen1992,Metzler2000and2004}, giving rise to anomalous behaviors that can be collectively described
as~\textit{strange kinetics}~\cite{Shlesinger1993}.

Equation~(\ref{eq-DiffusionCoeffScaling}) encompasses both the linear dependence of the mean squared displacement with
time~($\gamma = 1$) characteristic of the standard Brownian motion as well as nonlinear dependencies, slower~($\gamma < 1$)
and faster~($\gamma > 1$) than the classical case. The \textit{below} Browian motion regime is usually named sub-diffusion
or dispersive motion while the \textit{above} range is known as hyper or superdiffusion. In order to generalize the
classical treatment to consider anomalous diffusion, the Continuous Time Random Walk~(CTRW) model is often considered. The
random trajectory is viewed as the result of pairwise stochastically independent spatial and temporal increment events. When
both the variance of the spatial steps and the expectation value of the temporal increments are finite, CTRW is equivalent
to Brownian motion on large spatio-temporal scales~\cite{Metzler2000and2004} and yields ordinary diffusion. Nevertheless,
for the anomalous diffusion, a bifractional diffusion equation can be derived for the dynamics of $W\left( r,t\right)$. This
can be solved using the methods of fractional calculus to give

\begin{equation}
  \label{eq-Levy-Stable}
  W\left( r,t\right) \mathrel{\mathop{\sim }\limits_{r \, , t \rightarrow \infty }}
  \frac{1}{t^{1/\mu }}L_{\mu }\left( \frac{r}{t^{1/\mu }}\right)
  \mbox{,}
\end{equation}

where~$L_{\mu }$ is a one-sided L\'{e}vy stable law of index~$\mu$~\cite{Bouchaud1990}. This equation implies that the mean
squared displacement scales as

\begin{equation}
  \label{eq-Levy-power-law}
  p\left( r\right)
  \mathrel{\mathop{\sim }\limits_{r\rightarrow \infty }}
  r^{-(1+\mu )}
  \mbox{.}
\end{equation}

The particle's motion can be classified either as a L\'{e}vy flight, in the case that the \textit{time-of-flight} of each
jump in the trajectory is negligible, or as a L\'{e}vy walk, if the finite time to complete each jump must be taken into
account~\cite{Shlesinger1993,Metzler2000and2004}.

In this work we are interested in the stochastic theory of atomic resonance radiation migration under which the trajectories
are of the supperdiffusive type. The radiation transport is usually described in terms of a master equation, which has an
integro-differential form taking into account non-local effects~\cite{PaperI}. This is usually known as the
Holstein-Biberman equation, named after seminal contributions in the
1940s~\cite{Holstein1947,Molisch1998,Berberan-Santos1999}. It was shown, first using ad~hoc
arguments~\cite{Nunes-Pereira2004} and later demonstrated, that the motion of photons is superdiffusive. This was initially
demonstrated for Doppler and Lorentz spectral line shapes and recently generalized for any line shape~\cite{PaperI}. The
work was done in the limit of Complete Frequency Redistribution~(CFR), a situation in which the collision rate is
sufficiently high when compared to the lifetime of the excited state to destroy any correlation between photon absorption
and (re)emission events. This contribution aims to extend the previous work and discuss the influence of Partial Frequency
Redistribution~(PFR) effects in the stochastic description of the resonance radiation trajectories. We will show that the
overall asymptotic behavior of the trajectories is dictated by the redistribution in the rest frame of the atom, even if in
the lab rest frame memory effects persist in a finer scale. Specifically, we will show that the Doppler or Lorentz CFR
superdiffusive asymptotics are recovered for PFR if there is only Doppler or Doppler plus natural redistribution in the atom
frame. On the contrary, we show for the first time, that the complete coherence in the atom's rest frame will manifest
itself as a breakdown of the superdiffusive behavior in all cases and not only for the wing frequencies events.

In Sec.~II the Partial Frequency Redistribution~(PFR) effects are introduced along with the notation to be used. In Sec.~III
the single line Opacity Distribution Function~(ODF) under~PFR is defined. This is subsequently used to derive the the large
scale asymptotics of the jump length distribution in Sec.~IV and of the opacity distribution in Sec.~V. Section~VI makes a
connection between the superdiffusive character and the low opacity asymptotic behavior of the~ODF. In Sec.~VII numerical
results are presented for the ODFs and the conditional jump length probability distribution functions, and the influence of
frequency redistribution in both a small scale~(controlled by~PFR effects) as well as in a large scale~(corresponding to
either a~CFR superdiffusive or, in alternative, a diffusive) asymptotic are discussed. The main conclusions are summarized
at the end~(Sec.~VIII).

\section{PARTIAL FREQUENCY REDISTRIBUTION}

Resonance radiation trapping can be envisaged as a random flight with a jump size distribution dependent upon the spectral
line shape. If all the scattering~(absorption followed by reemission) events are coherent in frequency, the radiation
transport is described by a diffusion equation with a diffusion coefficient dependent on the mean free path of the
radiation. However, reabsorption-reemission events are inelastic, and a photon frequency redistribution in the lab reference
frame exists preventing the notion of a mean free path. Therefore, the topology of the excitation random trajectory has to
be related with the frequency redistribution of the emitted photons, since the jump length Probability Distribution
Function~(PDF) must consider all possible optical emission frequencies. This can be done as~\cite{Nunes-Pereira2004}

\begin{eqnarray}
  \label{eq-PDFprPFR}
  p\left( r\right) & = & \int_{-\infty }^{+\infty }\Theta \left( x\right) \,p\left( r|x\right) \,\dif x \nonumber \\
  & = & \int_{-\infty }^{+\infty }\Theta \left( x\right) \Phi \left( x\right) \me ^{-\Phi \left( x\right) r}\,\dif x
  \mbox{,}
\end{eqnarray}

where~$x$ is a frequency difference to the center-of-line frequency~$x = \frac{\nu -\nu _{0}}{W}$, properly normalized by
dividing by the corresponding width parameter~(given for the Doppler and Lorentz spectral profiles as~$W_{Dop}=v_{mp}\,
\frac{\nu _{0}}{c}$ and~$W_{Lor}=\Gamma /4\pi $, respectively, with~$v_{mp}$ the most probable thermal velocity and~$\Gamma
$ the radiative rate constant), $\Theta\left( x\right)$ is the emission spectrum lineshape, $p\left( r|x\right)$ the photon
exponential~(Beer-Lambert) jump distribution, and $\Phi \left( x\right)$ is the normalized absorption lineshape of the
resonance line(~so that~$\int_{-\infty }^{+\infty }\Phi \left( x\right) \,\dif x $). The dimensionless distance
$r=\frac{k_{0}}{\Phi \left( 0\right)}$ defines an \textit{opacity} or \textit{optical density} scale from the center-of-line
optical depth~$k_{0}$ and the center-of-line absorption coefficient~$\Phi \left( 0\right)$. For a sufficiently high number
of collisions during the lifetime of excited atoms, the emitted photon has a frequency completely uncorrelated with the
absorption frequency and so the radiation is completely redistributed over the entire spectral line. This corresponds to the
Complete Frequency Redistribution~(CFR) case, for which the emission and absorption line shapes coincide and the jump length
PDF is independent of the past history. For strong resonance lines with short lifetimes the frequency redistribution is
partial since only a few collisions occur before emission, and consequently the stochastic nature of the distribution must
be considered. Partial Frequency Redistribution~(PFR) effects are important in an astrophysical context, as is the case of
the scattering of Ly$\alpha $ radiation in optically thick nebulae~\cite{Frisch1980} and the Ca~I resonance line in the
solar spectrum~\cite{Frisch1996}, as well as in lab scale atomic vapors, notably the $185$~nm mercury line in low-pressure
lighting discharges (which can account for as many as $10\%$ of the overall flux in a typical T12 fluorescence
lamp)~\cite{Lister2004} and the $147$ and $129$~nm Xe VUV radiation used in Plasma Display Pannels~(PDPs)
applications~\cite{Kushner_and_PDPs,Anderson1995}. For both mercury and xenon, the natural lifetimes of the corresponding
excited states are less than about $3$~ns and therefore only at very high densities is the collision rate high enough to
approach CFR conditions.

For a two-level model in the CFR limit, both absorption and emission spectra lineshapes can be described by
Doppler~\textendash ~$\Phi _{Dop}\left( x\right) = \frac{1}{\sqrt{\mpi}}\me ^{-x^{2}}$~\textendash , Lorentz~\textendash
~$\Phi_{Lor}\left( x\right) =\frac{1}{\mpi}\frac{1}{1+x^{2}}$~\textendash , or Voigt~\textendash~$\Phi
_{Voigt}\left(x\right) =\frac{a}{\mpi ^{3/2}}\int_{-\infty }^{+\infty }\frac{\me ^{-u^{2}}} {a^{2}+\left( x-u\right)
^{2}}\,\dif u$~\textendash ~spectral distributions, in which $x$ is a normalized~(with the Doppler width) difference to the
center of line frequency and $a$ is the Voigt characteristic width, a ratio of the Lorentz to the Doppler widths defined
previously. The study of PFR was pioneered in the 1940s in an astrophysical context and we will use the notation introduced
by Hummer in the early 1960s~\cite{Hummer1962}. PFR is described by a redistribution $R$ function which is the joint
probability of absorption of a $x'$ photon \textit{and} reemission of a $x$~frequency photon. This redistribution function
is derived for a thermal vapor, assuming a Maxwell-Boltzmann velocity distribution for ground state atoms and further
imposing that the atoms velocity is unchanged upon excitation~(assumption not valid for ultracold vapors).

We will further consider unpolarized radiation~(scalar redistribution function), use the angle averaged redistribution for
an isotropic angular distribution reemission and discuss the three most important cases for a ground state absorbing state:
(i)~pure Doppler broadening~($R_{I}$), (ii)~combined natural and Doppler broadening~($R_{II}$) and (iii)~combined Doppler,
natural and collisional broadening~($R_{III}$)~\cite{Mihalas1978,Molisch1998,Hubeny1985A,Frisch1980} . For the $R_{I}$ case,
both levels are infinitely sharp and there is pure coherent scattering in the Atom's Rest Frame~(ARF). This is unrealistic
because both the lower and upper levels of the optical transition are broadened to some extent by radiation damping and/or
collisions. It is nevertheless useful since it allows the separation of the Doppler shift from the other broadening
mechanisms. The physical picture for $R_{II}$ is a line with an infinitely sharp lower level and an upper level broadened
only by radiation damping. In the~ARF the absorption is Lorentzian and scattering is completely (frequency) coherent. This
type of scattering is important for resonance lines in low density media where the collisions do not perturb the excited
levels. $R_{III}$ applies whenever the upper state is broadened by radiation and collision damping in the limit where
collisions are frequent enough to cause complete frequency redistribution in the ARF. Both absorption and reemission
profiles are Lorentzian in the atom's frame with a total width equal to the sum of the radiative plus
collisional~(uncorrelated) characteristic widths. It is well known that complete redistribution in the ARF does not imply
the same quantitative behavior in the lab frame used to describe radiation transport. Under the previously mentioned
assumptions, the lab frame \textit{joint} redistribution functions are given by~\cite{Mihalas1978,Cannon1985}

\begin{equation}
  \label{eq-RI}
  R_{I}\left( x^{\prime };x\right) =\frac{1}{2}
  \erfc \left( \overline{x} \right)
  \mbox{,}
\end{equation}

\begin{eqnarray}
  \label{eq-RII}
  R_{II}\left( x^{\prime };x\right) & = & \frac{1}{\pi ^{3/2}}
  \int_{\frac{1}{2}\left\vert x-x^{\prime }\right\vert }^{+\infty }
  e^{-u^{2}} \times \nonumber \\
  & & \times \left\{ \arctan \left[ \frac{\underline{x}+u}{a}\right]
  -\arctan \left[ \frac{\overline{x}-u}{a}\right] \right\} \,\dif u
  \mbox{,}
\end{eqnarray}

and

\begin{eqnarray}
  \label{eq-RIII}
  R_{III}\left( x^{\prime };x\right) & = & \frac{1}{\pi ^{5/2}}
  \int_{0}^{+\infty } e^{-u^{2}} \times \nonumber \\
  & & \times \left\{ \arctan \left[ \frac{x^{\prime }+u}{a}\right]
  -\arctan \left[ \frac{x^{\prime }-u}{a}\right] \right\} \\
  & & \times \left\{ \arctan \left[ \frac{x+u}{a}\right]
  -\arctan \left[ \frac{x-u}{a}\right] \right\}
  \,\dif u \nonumber
  \mbox{,}
\end{eqnarray}

where $a$ is the total width, $\erfc$ is the complementary error function, and $\underline{x}$ and $\overline{x}$ stand for
$\underline{x} \equiv \min \left( \left\vert x^{\prime }\right\vert ,\left\vert x\right\vert \right) $ and $ \overline{x}
\equiv \max \left( \left\vert x^{\prime }\right\vert ,\left\vert x\right\vert \right) $.

In Eq.~(\ref{eq-PDFprPFR}) we need to use the (re)emission PDF. For radiation transport in thermal vapors we can assume a
CFR absorption profile which corresponds to the ensemble average over the Maxwellian velocity distribution. The reemission
distribution comes from the joint redistribution probability and, according to Bayes rule, is given by

\begin{equation}
  \label{eq-Bayes}
  \Theta \left( x|x^{\prime }\right) =\frac{R\left( x^{\prime };x\right) }{\Phi \left( x^{\prime }\right) }
  \mbox{,}
\end{equation}

and it is this conditional (re)emission PDF that should be used. $\Phi \left( x^{\prime }\right) $ is the absorption
spectrum PDF which should be either the pure Doppler distribution~($R_{I}$) or the Voigt profile~($R_{II}$ and $R_{III}$).

\begin{figure}
  \includegraphics[width=10cm,keepaspectratio=true]{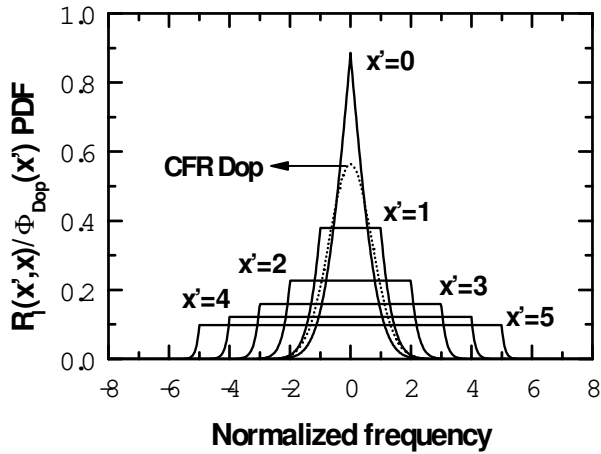}
    \caption{\label{Fig-RI}
    Conditional reemission Probability Distribution Functions~(PDF) of Partial Redistribution function~$R_{I}$ for several
    previous absorption frequencies~$x^{\prime}$. The corresponding Doppler Complete Redistribution is also shown.}
\end{figure}

\begin{figure}
  \includegraphics[width=10cm,keepaspectratio=true]{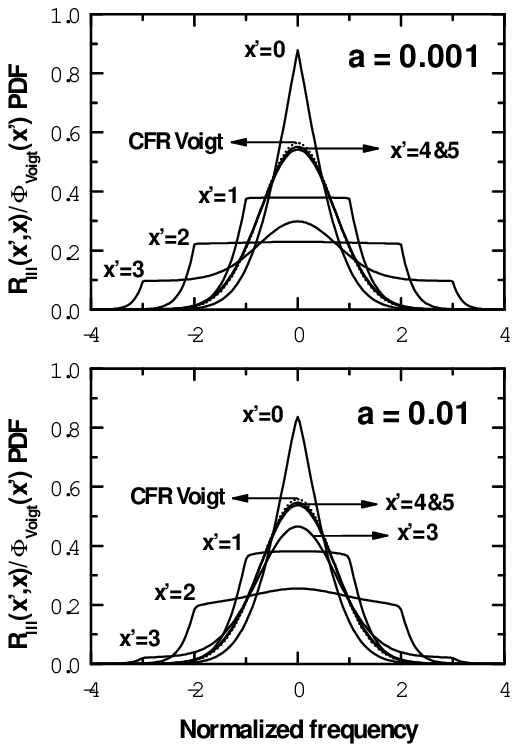}
  \caption{\label{Fig-RII}
    Conditional reemission Probability Distribution Functions~(PDF) of Partial Redistribution function~$R_{III}$ for several
    previous absorption frequencies~$x^{\prime}$ and Voigt width parameters~$a$. The corresponding Voigt
    Complete Redistribution is also shown.}
\end{figure}

\begin{figure}
  \includegraphics[width=10cm,keepaspectratio=true]{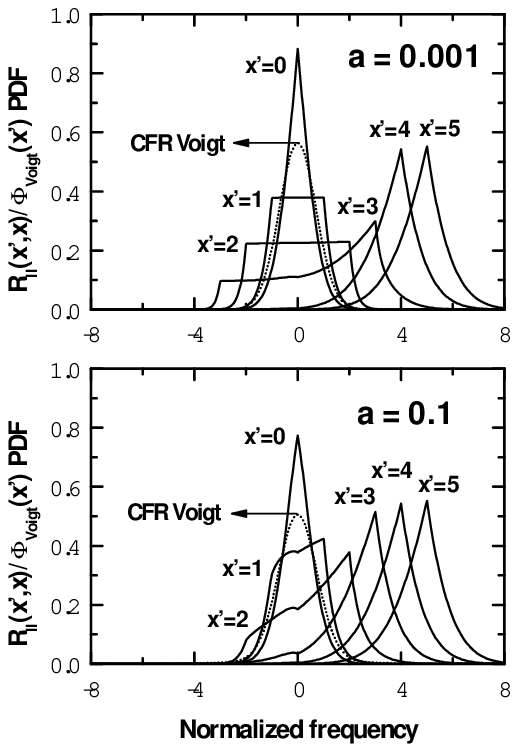}
  \caption{\label{Fig-RIII}
    Conditional reemission Probability Distribution Functions~(PDF) of Partial Redistribution function~$R_{II}$ for several
    previous absorption frequencies~$x^{\prime}$ and Voigt width parameters~$a$. The corresponding Voigt
    Complete Redistribution is also shown.}
\end{figure}

Figures~\ref{Fig-RI} to~\ref{Fig-RIII} show the spectral distributions calculated from the previous equations. Both $R_{I}$
and $R_{III}$ distributions are symmetric, while $R_{II}$ is asymmetric. In all cases, a complete redistribution in the~ARF
does not imply CFR in the lab frame, the degree of redistribution in the lab frame being always bigger for core radiation
and for the $R_{I}$ and $R_{III}$ cases. The complete coherence in the atom's rest frame for the $R_{II}$ model is not
complete in the lab frame. However, the coherence degree is more retained for wing photons and higher~$a$ values~(bigger
importance of natural coherent broadening compared to Doppler). On the other hand, for the $R_{III}$ distribution, the
increase of the $a$ value implies higher redistribution in the core~(bigger importance of collisional redistributed
broadening). We will discuss now the influence of the redistribution between absorption and reemission in the asymptotic
superdiffusive character of photon migration under PFR conditions. For this purpose we extend the concept of a single line
opacity distribution function for incomplete redistribution.

\section{LINE OPACITY DISTRIBUTION FUNCTION UNDER PARTIAL REDISTRIBUTION}

To study the superdiffusive behavior of resonance radiation trapping it is convenient to discuss the influence of the line
shape in the Opacity (probability) Distribution Function~(ODF). For that we will use a line opacity scale, defined from the
normalized absorption line shape as $k\equiv \Phi \left( x\right) $. This line opacity scale will thus have no reference to
the actual system size or number density and will be useful only for infinite media. This is the one chosen to discuss
the~ODF. The monochromatic line opacity along a given pathlength~$l$, for homogeneously distributed particles, is~$k\left(
x\right) =n\sigma _{0} l \Phi \left( x\right) / \Phi \left( 0\right) = k_{0}\Phi \left( x\right) / \Phi \left( 0\right) =
\Phi \left( x\right) r$ and is therefore proportional to the number density~$n$~(or the pressure) and to the center-of-line
opacity~$k_{0} =n\sigma _{0} l$. $\sigma _{0}$ is the center-of-line absorption cross section as usual. The discussion of
the spectral line profile influence on the jump length distribution will use an overall (for the whole line) dimensionless
opacity or optical density scale defined as~$r = \frac{k_{0}}{\Phi \left( 0\right)}$ so that~$r=\int_{-\infty }^{+\infty
}k\left( x\right) \,\dif x$. This is done since any given jump distance must reflect the whole of the line shape and not
just a given monochromatic opacity.

We will now generalize the basic procedure used previously~\cite{PaperI} for the complete frequency distribution to cover
the partial frequency distribution~(PFR) case. We will start with the jump length PDF by rewriting Eq.~(\ref{eq-PDFprPFR})
making explicit the conditional dependence in the absorption frequency:

\begin{equation}
  \label{eq-PDFprPFR-xline}
  p\left( r | x^{\prime }\right)
  =\int_{-\infty }^{+\infty }\Theta \left( x | x^{\prime }\right) \Phi \left( x\right) \me ^{-\Phi \left( x\right) r}\,\dif x
  \mbox{.}
\end{equation}

Since photon migration by reabsorption is associated with exponential jump size PDFs, it is desirable to consider the
general $p\left( r | x^{\prime }\right)$ as a linear combination of exponential densities given by

\begin{equation}
  \label{eq-H-xline}
  p\left( r | x^{\prime }\right)
  =\int_{0}^{+\infty }H\left( k|x^{\prime }\right) \left( k\,e^{-k\,r}\right)\,\dif k
  \mbox{,}
\end{equation}

where $H\left( k|x^{\prime }\right)$ is the PDF of effective line opacities. The single line opacity at a given frequency,
$x$, is given by $k\left( x\right) \equiv  \Phi \left( x\right) $. As was shown before for the CFR case, $k\,H\left(
k|x^{\prime }\right)$ is the inverse Laplace transform of $p\left( r | x^{\prime }\right)$. The inversion is analytical and
can be performed using the real inversion form of the Laplace transform (with the parameter
$c=0$)~\cite{Berberan-Santos2005},

\begin{equation}
  k\, H\left( k|x^{\prime }\right) =\frac{1}{\pi }
  \int_{0}^{+\infty }\left\{ \Real\left[ p\left( i\omega \right) \right] \cos \left(k\omega \right)
  -\Imag\left[ p\left( i\omega \right) \right] \sin \left( k\omega \right) \right\} \,\dif \omega
  \mbox{,}
\end{equation}

which gives~\cite{PaperI}

\begin{equation}
  H\left( k|x^{\prime }\right) =\frac{1}{k}
  \int_{-\infty }^{+\infty }\Theta \left( x|x^{\prime }\right) \, \Phi \left( x\right) \,
  \delta \left[ k-\Phi \left( x\right) \right] \, \dif x
  \mbox{.}
\end{equation}

This can be further simplified by making the change of variable~$y = \Phi \left( x\right)$ and decomposing the integration
into positive and negative~$x$ frequencies,

\begin{equation}
  \label{eq-H-notfinal}
  H\left( k|x^{\prime }\right) = -\frac{1}{k}
  \int_{0}^{\Phi \left( 0\right) }\left[ \theta _{+}\left( k\right) +\theta _{-}\left( k\right) \right] \, y\,
  \Psi ^{\prime }\left( y\right) \delta \left[ k-y\right] \, \dif y
  \mbox{,}
\end{equation}

since it is assumed that $\Phi \left( x\right)$ is always nonnegative, and that the transformation has opposite signs for
positive and negative frequencies. $\Psi \left( y\right)$ is the inverse function of~$\Phi \left( x\right)$, $\Psi \left(
y\right) = \Phi ^{-1}\left( y\right)$. The integration in $y$ only runs to the maximum value of the absorption spectrum~(the
\textit{center-of-line} absorption coefficient, $\Phi \left( 0\right) $) and $ \theta _{+}\left( k\right)$ and $ \theta
_{-}\left( k\right)$ represent the value of the (conditional) probabilities for emission for \textit{positive} and
\textit{negative} frequencies, implicitly given from the opacity~$k$ through the absorption spectrum. For any particular
opacity~$k$, there are two (symmetrical) frequencies, $+x$ and $-x$, that give rise to~$k$ defined by~$\Phi \left( x \right)
= k$~(symmetric absorption). These frequencies do not necessarily have the same reemission probability. One therefore can
define $\theta _{+}\left( k\right) = \Theta \left( x|x^{\prime }\right) $ and $\theta _{-}\left( k\right) = \Theta \left(
-x|x^{\prime }\right) $ and finally rewrite Eq.~(\ref{eq-H-notfinal}) as

\begin{equation}
  \label{eq-H-final}
  H\left( k|x^{\prime }\right) = \left\{
  \begin{array}{c@{\quad \mbox{if} \quad}l}
    -\left[ \theta _{+}\left( k\right) +\theta _{-}\left( k\right) \right]
    \Psi ^{\prime }\left( k\right) & k<\Phi \left( 0\right) \\
    0 & k\geq \Phi \left( 0\right)
  \end{array}
  \right.
  \mbox{.}
\end{equation}

This is the general workhorse that will be used in the following sections to discuss the influence of PFR in resonance
radiation reabsorption and compare it with the CFR case~\cite{PaperI}. We will show how the redistribution of the emission
into the wings of the absorption spectrum implies a superdiffusive behavior characterized by a very broad~(infinite mean)
jump size distribution. In addition, we are able to demonstrate that, if redistribution into the tails of the absorption is
prevented due to PFR memory effects, the superdiffusive character breaks down, and this can be seen in a cut-off in the PDF
of the opacity at small line opacity values that is translated into a corresponding cut-off of the higher jump lengths.

Two of the PFR cases defined before ($R_{I}$ and $R_{III}$) have symmetric reemission PDFs (see Figs.~1 and 2), which allow
us to write,

\begin{equation}
  \label{eq-H-symmetrical}
  H\left( k|x^{\prime }\right) = \left\{
  \begin{array}{c@{\quad \mbox{if} \quad}l}
    - 2 \theta _{+}\left( k\right) \Psi ^{\prime }\left( k\right) & k<\Phi \left( 0\right) \\
    0 & k\geq \Phi \left( 0\right)
  \end{array}
  \right.
  \mbox{.}
\end{equation}

For Complete Frequency Redistribution this gives

\begin{equation}
  \label{eq-H-symmetrical-CFR}
  H\left( k|x^{\prime }\right) = \left\{
  \begin{array}{c@{\quad \mbox{if} \quad}l}
    - 2 k \, \Psi ^{\prime }\left( k\right) & k<\Phi \left( 0\right) \\
    0 & k\geq \Phi \left( 0\right)
  \end{array}
  \right.
  \mbox{,}
\end{equation}

which is simply the result obtained before~\cite{PaperI}.

\section{ASYMPTOTIC BEHAVIOR OF THE JUMP DISTRIBUTION}

Using Eq.~(\ref{eq-H-final}), Eq.~(\ref{eq-H-xline}) can be rewritten as

\begin{equation}
  p\left( r | x^{\prime }\right)
  =-\int_{0}^{\Phi \left( 0\right) }
  \left[ \theta _{+}\left( k\right) +\theta _{-}\left( k\right) \right] \,
  \Psi ^{\prime }\left( k\right) \left( k \, e^{-kr}\right) \, \dif k
  \mbox{.}
\end{equation}

The asymptotic behavior is easily found by performing the change of variable $y = k \, r$,

\begin{equation}
  p\left( r | x^{\prime }\right)
  =- \frac{1}{r^{2}} \int_{0}^{\Phi \left( 0\right) \, r}
  \left[ \theta _{+}\left( \frac{y}{r}\right) +\theta _{-}\left( \frac{y}{r}\right) \right] \,
  \Psi ^{\prime }\left( \frac{y}{r}\right) \left( y \, e^{-y}\right) \, \dif y
  \mbox{,}
\end{equation}

which gives

\begin{equation}
  \label{eq-asymptotic-notfinal}
  p\left( r | x^{\prime }\right)
  \mathrel{\mathop{\sim }\limits_{r\rightarrow \infty }}
  r^{-2} \int_{0}^{+\infty }
  \left[ \theta _{+}\left( \frac{y}{r}\right) +\theta _{-}\left( \frac{y}{r}\right) \right] \,
  \Psi ^{\prime }\left( \frac{y}{r}\right) \left( y \, e^{-y}\right) \, \dif y
  \mbox{.}
\end{equation}

For PFR conditions, we have to consider both the absorption as well as the conditional reemission asymptotics. For the
absorption spectrum, if the line shape function can be given by a power-law asymptotic in the
wings~\cite{Nunes-Pereira2004,PaperI}, then one has

\begin{equation}
  \label{eq-asymptotic-a}
  k\left( x\right) \equiv \Phi \left( x\right)
  \mathrel{\mathop{\sim }\limits_{x\rightarrow \infty }}
  x^{-p_{a}} \quad \mbox{with} \quad (p_{a} > 1)
  \mbox{,}
\end{equation}

where~$a$ stands for absorption.

The inverse function is

\begin{equation}
  \label{eq-asymptotic-a-inverse}
  \Psi \left( k\right)
  \mathrel{\mathop{\sim }\limits_{k\rightarrow 0 }}
  k^{-\frac{1}{p_{a}}}
  \mbox{,}
\end{equation}

which can be written, as

\begin{equation}
  \Psi ^{\prime }\left( \frac{y}{r}\right)
  \mathrel{\mathop{\sim }\limits_{r\rightarrow \infty }}
  - \frac{r}{p_{a} y} y^{-\frac{1}{p_{a}}} r^{\frac{1}{p_{a}}}
  \mbox{.}
\end{equation}

If the asymptotic of the conditional reemission is similar to that of absorption~(but possibly with a different parameter
value), then

\begin{equation}
  \label{eq-asymptotic-e}
  \Theta \left( x|x^{\prime }\right)
  \mathrel{\mathop{\sim }\limits_{x\rightarrow \infty }}
  x^{-p_{e}} \quad \mbox{with} \quad (p_{e} > 1)
  \mbox{,}
\end{equation}

where~$e$ stands for emission. Therefore, both~$\theta _{+}$ and $\theta _{-}$ scale as

\begin{equation}
  \label{eq-asymptotic-e-inverse}
  \theta _{+} \left( k\right) \sim \theta _{-}\left( k\right)
  \mathrel{\mathop{\sim }\limits_{k\rightarrow 0 }}
  k^{\frac{p_{e}}{p_{a}}}
  \mbox{.}
\end{equation}

Finally, Eq.~(\ref{eq-asymptotic-notfinal}) can be rewritten as

\begin{eqnarray}
  p\left( r | x^{\prime }\right)
  & \mathrel{\mathop{\sim }\limits_{r\rightarrow \infty }} & r^{-2} \int_{0}^{+\infty }
  \left( \frac{y}{r}\right) ^{p_{e}/p_{a}} \left[ -\left( \frac{r}{p_{a}y}\right) y^{-1/p_{a}}r^{1/p_{a}}\right]
  \left( ye^{-y}\right) \, \dif y \nonumber \\
  & = & \Gamma \left( 1+\frac{p_{e}}{p_{a}}-\frac{1}{p_{a}}\right) \,
  r^{\frac{1}{p_{a}}-\left( 1+\frac{p_{e}}{p_{a}}\right) }
  \mbox{,}
\end{eqnarray}

and therefore,

\begin{equation}
  \label{eq-asymptotic-final}
  p\left( r | x^{\prime }\right)
  \mathrel{\mathop{\sim }\limits_{r\rightarrow \infty }}
  r^{\frac{1}{p_{a}}-\left( 1+\frac{p_{e}}{p_{a}}\right) }
  \mbox{.}
\end{equation}

Comparing the previous equation with the L\'{e}vy flight asymptotic in Eq.~(\ref{eq-Levy-power-law}), gives

\begin{equation}
  \label{eq-Levy-PFR}
  \mu = \frac{1}{p_{a}} \left( p_{e} - 1 \right)
  \mbox{.}
\end{equation}

If the asymptotic regime is the same for both absorption and PFR reemission, $p = p_{a} = p_{e}$, and the jump size
distribution has the asymptotic

\begin{equation}
  \label{eq-asymptotic-CFR}
  p\left( r | x^{\prime }\right)
  \mathrel{\mathop{\sim }\limits_{r\rightarrow \infty }}
  r^{-(2-\frac{1}{p}) }
  \mbox{,}
\end{equation}

which corresponds to the CFR form obtained before~\cite{PaperI}. So, the CFR asymptotic is recovered even in PFR conditions,
as long as the asymptotic forms of both the absorption spectrum and the conditional reemission are the same. The L\'{e}vy
parameter is also the one obtained in CFR conditions,

\begin{equation}
  \label{eq-Levy-CFR}
  \mu = 1- \frac{1}{p}
  \mbox{.}
\end{equation}

\section{ASYMPTOTIC BEHAVIOR OF THE OPACITY DISTRIBUTION}

We can obtain the same conclusions from the spectral and from the ODF PDFs asymptotics. Indeed, the small opacity scaling
law can be easily obtained from the previous equations. From Eqs.~(\ref{eq-asymptotic-a})
and~(\ref{eq-asymptotic-a-inverse}), one obtains

\begin{equation}
  \Psi ^{\prime }\left( k \right)
  \mathrel{\mathop{\sim }\limits_{k\rightarrow 0}}
  - \frac{1}{p_{a}} k^{-\frac{1}{p_{a}}-1}
  \mbox{.}
\end{equation}

Using the reemission asymptotic of Eq.~(\ref{eq-asymptotic-e-inverse}), the ODF small opacity limit is

\begin{equation}
  \label{eq-asymptotic-H}
  H\left( k|x^{\prime }\right)
  \mathrel{\mathop{\sim }\limits_{k\rightarrow 0}}
  - \frac{1}{p_{a}} k^{- \left[ \frac{1}{p_{a}}+\left( 1-\frac{p_{e}}{p_{a}}\right) \right]}
  \mbox{,}
\end{equation}

which, for absorption and reemission spectral distributions with the same asymptotic character~$p = p_{a} = p_{e}$, gives

\begin{equation}
  H\left( k|x^{\prime }\right)
  \label{eq-asymptotic-H-linear}
  \mathrel{\mathop{\sim }\limits_{k\rightarrow 0}}
  - \frac{1}{p} k^{- \frac{1}{p}}
  \mbox{.}
\end{equation}

The last two equations show that the ODF asymptotic behavior can be easily identified from a $log$-$log$ plot as

\begin{equation}
  \log H\left( k|x^{\prime }\right)
  \mathrel{\mathop{\sim }\limits_{k\rightarrow 0}}
  - \left[ \frac{1}{p_{a}}+\left( 1-\frac{p_{e}}{p_{a}}\right) \right] \log k
  \mbox{,}
\end{equation}

and

\begin{equation}
  \label{eq-asymptotic-H-CFR}
  \log H\left( k|x^{\prime }\right)
  \mathrel{\mathop{\sim }\limits_{k\rightarrow 0}}
  - \frac{1}{p} \log k
  \mbox{.}
\end{equation}

The cases of CFR Doppler and Lorentz resonance line radiation transfer give analytical ODF expressions and were considered
in previous works~\cite{PaperI,Nunes-Pereira2004}. For the Doppler case, one has~$\mu \simeq 1$ and thus
Eq.~(\ref{eq-asymptotic-a}) is only approximately valid and the dependence of the ODF over the opacity values is extremely
weak. However, for Lorentz, the asymptotic in Eq.~(\ref{eq-asymptotic-a}) is exact with~$p = 2$. The ODF is given
by~\cite{PaperI}

\begin{equation}
  H_{Lor}\left( k\right) =\frac{\Phi \left( 0\right) }{k}\left( \frac{\Phi \left( 0\right) }{k}-1\right) ^{-1/2}
  \mbox{,}
\end{equation}

which has an exact asymptotic of the form given by Eq.~(\ref{eq-asymptotic-H-CFR}).

\section{SUPERDIFFUSIVE BEHAVIOR FOR RESONANCE LINE TRAPPING}

Although the superdiffusive behavior is well established from the asymptotics derived in the previous two sections, a
simpler more informative, alternative approach is possible. Equation~(\ref{eq-H-xline}) can be written in Laplace space as

\begin{equation}
  \label{eq-H-xline-Lap}
  \hat{p}\left( u | x^{\prime }\right)
  =\int_{0}^{\Phi \left( 0\right) } \frac{ k\,H\left( k|x^{\prime }\right) } { u+k } \,\dif k
  \mbox{,}
\end{equation}

and since the moments of the jump size distribution can be recast from the $m$th~derivatives of this Laplace transform as~$
\left. \hat{p}^{\left( m\right)}\left( u | x^{\prime }\right) \right\vert _{u=0} = \left( -1\right)^{m} \, \left\langle
r^{m}\right\rangle $, one obtains

\begin{equation}
  \label{eq-moments}
  \left\langle r^{m}\right\rangle
  = m! \, \int_{0}^{\Phi \left( 0\right) } \frac{ H\left( k|x^{\prime }\right) } { k^{m} } \,\dif k
  = m! \, \left\langle k^{-m}\right\rangle
  \mbox{.}
\end{equation}

The integration only goes up to the upper limit of the center-of-line opacity. The first moment gives~$\left\langle
r\right\rangle = \int_{0}^{\Phi \left( 0\right) } \frac{ H\left( k|x^{\prime }\right) } { k } \,\dif k$ which shows that,
unless~$H\left( k|x^{\prime }\right)$ vanishes as~$k \rightarrow 0$, an infinite mean jump size is obtained.

If one considers now the asymptotic of Eq.~(\ref{eq-asymptotic-H-linear}), one can conclude that~$\left\langle
r\right\rangle \propto \underset{k\rightarrow 0^{+}}{\lim }\frac{1}{k^{1/p}}$ which is finite only if~$1/p \leq 0$. But,
since~$p \geq 1$ by construction, one can finally conclude that the mean jump size is infinite irrespective of the actual
value of the~$p$ parameter.

\section{NUMERICAL RESULTS}

\begin{figure}
  \includegraphics[width=10cm,keepaspectratio=true]{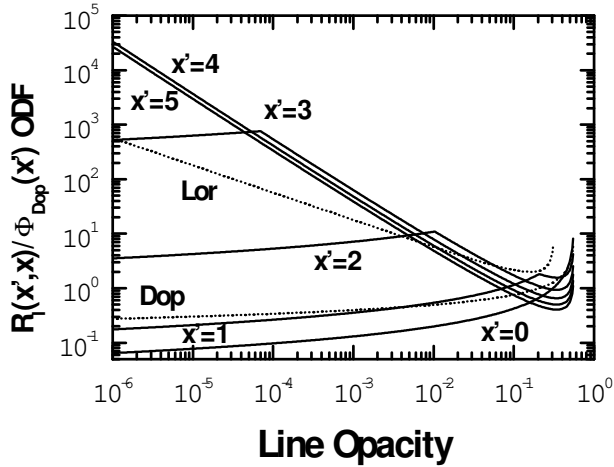}
  \caption{\label{Fig-H-RI}
    Conditional Opacity Distribution Function~(ODF) for Partial Redistribution function~$R_{I}$ for several
    previous absorption frequencies~$x^{\prime}$. The Complete Redistribution limiting Doppler and Lorentz cases are also shown.}
\end{figure}

\begin{figure}
  \includegraphics[width=10cm,keepaspectratio=true]{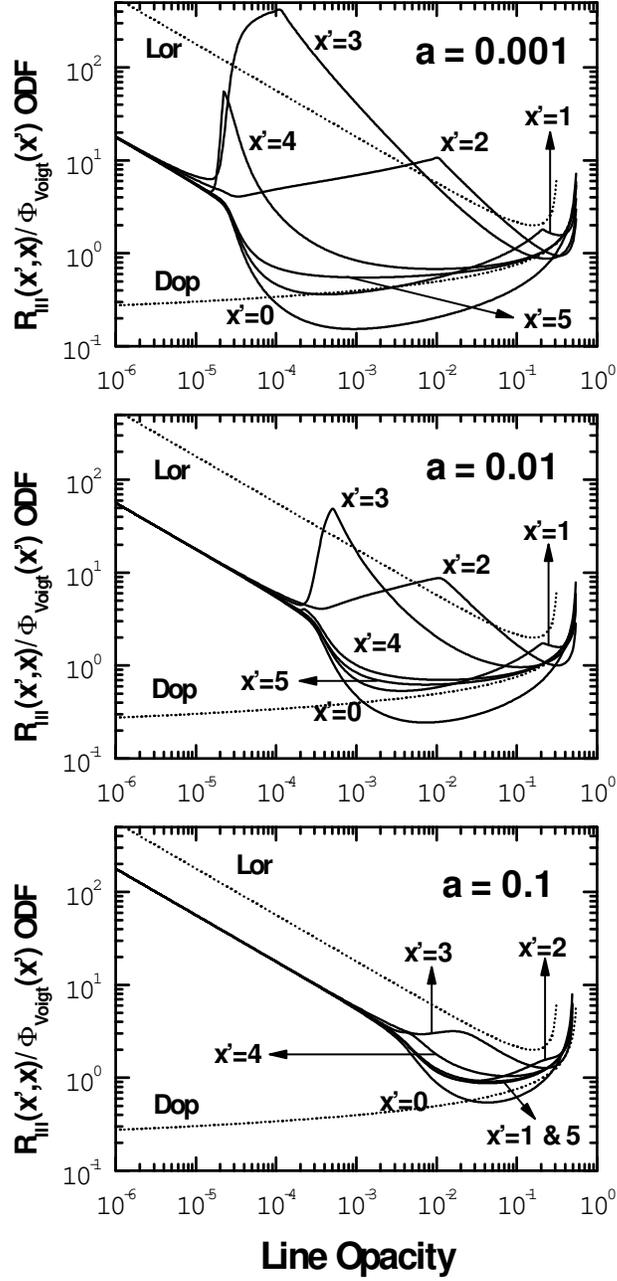}
  \caption{\label{Fig-H-RIII}
    Conditional Opacity Distribution Function~(ODF) for Partial Redistribution function~$R_{III}$ for several
    previous absorption frequencies~$x^{\prime}$ and Voigt width parameters~$a$. The Complete Redistribution limiting
    Doppler and Lorentz cases are also shown.}
\end{figure}

\begin{figure}
  \includegraphics[width=10cm,keepaspectratio=true]{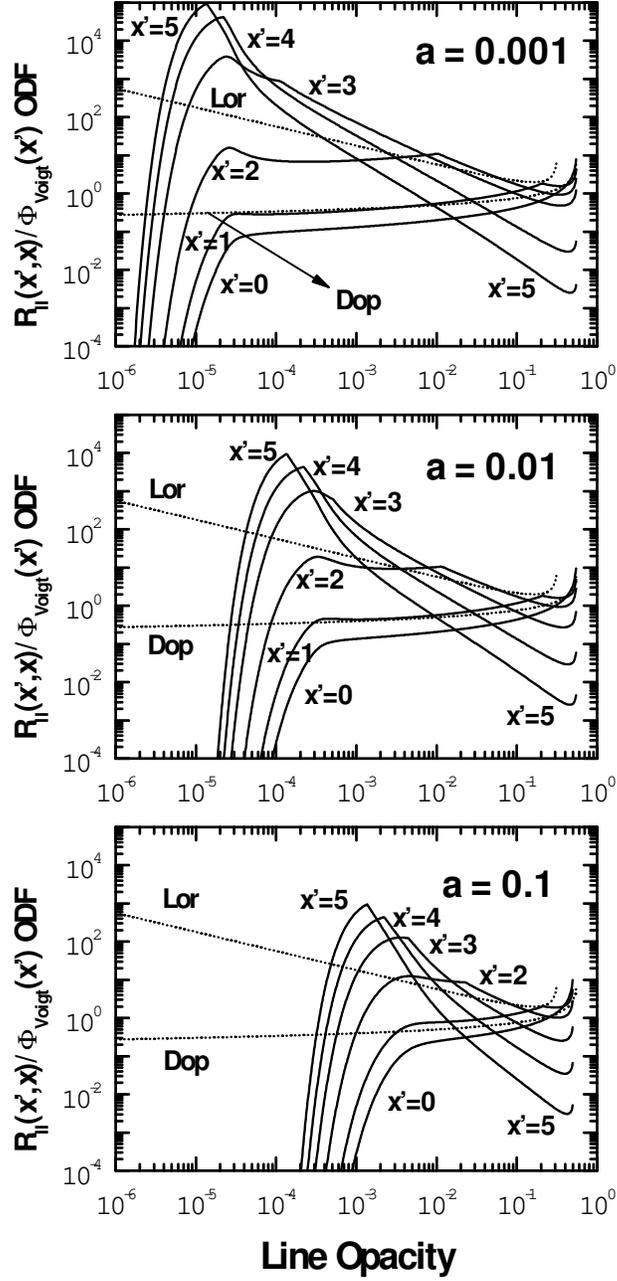}
  \caption{\label{Fig-H-RII}
    Conditional Opacity Distribution Function~(ODF) for Partial Redistribution function~$R_{II}$ for several
    previous absorption frequencies~$x^{\prime}$ and Voigt width parameters~$a$. The Complete Redistribution
    limiting Doppler and Lorentz cases are also shown.}
\end{figure}

Figures~\ref{Fig-H-RI} to~\ref{Fig-H-RII} give the conditional ODFs for the three PFR redistribution mechanisms considered
in this work alongside with the corresponding CFR Doppler and Lorentz limiting cases. The results were computed from
Eqs.~(\ref{eq-H-final}) and~(\ref{eq-H-symmetrical}). The CFR ODFs allow one to quantify very easily the cause of the
superdiffusive behavior for resonance line radiation trapping; for very small opacities, the ODF either decreases very
slowly~(Doppler) or even increases in importance~(Lorentz) with a decrease in opacity values. It does not vanish in the
limit of~$k\rightarrow 0$ and therefore superdiffusion sets in~(see previous section). From Eq.~(\ref{eq-H-xline}), one can
see that the slower the decrease in the ODF for smaller opacity values, the more important the higher jump sizes. For
Doppler, Lorentz~(and Voigt) CFR trapping, the ODF variation for small~$k$ values is either increasing or so slowly
decreasing as to render even the first moment of the jump size distribution infinite~\cite{Nunes-Pereira2004,PaperI}. The
CFR Voigt case is a self-affine multifractal with two different characteristic L\'{e}vy flight~$\mu$'s parameters, a~$\mu
\simeq 1$ corresponding to the Doppler-like core radiation and a~$\mu = 1/2$ for Lorentz-like wing photons. The overall
excitation random walk topology is controlled by the bigger Lorentz scale~(if not truncated by the vapor confining
boundaries)~\cite{Nunes-Pereira2004}. Therefore, in the present case of PFR effects in an infinite medium, and even for the
Voigt Maxwellian absorption profile, we need only concern ourselves with the PFR deviation from the most extreme Lorentz CFR
asymptotic.

Figure~\ref{Fig-H-RI} shows that the superdiffusive character of~$R_{I}$ partial redistribuion is similar to the one for the
corresponding Doppler CFR case. There is a transition between two regimes:~(i)~the central core constant
redistribution~(compare Fig.~\ref{Fig-RI}), characterized by a faster than Lorentz increase of opacity PDF for small opacity
values, and~(ii)~the wings Doppler-like redistribution, with a Doppler-like~(albeit with a slower convergence) asymptotic.
The transition between the two regimes depends upon the actual value of the previous absorption
frequency~$x^{\prime}$~(for~$x^{\prime} = 4$ and~$5$, the transition still occurs but at opacity values smaller than the
ones represented in Fig.~\ref{Fig-H-RI}).

The~$R_{I}$ case is useful because it allows one to single out the pure Doppler effect on the redistribution functions. The
more realistic cases are described by the~$R_{II}$ or~$R_{III}$ distributions. Figure~\ref{Fig-H-RIII} show the conditional
ODFs for the~$R_{III}$ redistribution for representative values of the~$a$~parameter found in experimental conditions. The
PFR random walk has a superdiffusive character with the same asymptotic of the CFR Lorentz. For the~$R_{III}$ combined
Doppler, natural and collisional broadening, there is complete (Lorentzian) redistribution in the atom's rest frame, but not
in the lab frame. One can distinguish mainly two regimes, one for very small line opacities and the other for opacities
approaching the center-of-line value, the exact transition between the two depending upon the~$a$ parameter. For very small
line opacities, the actual superdiffusive regime corresponds to the CFR Lorentz case. For higher line opacities, the
redistribution is similar to~$R_{I}$; an initial step increase in ODF PDF with decreasing opacity, followed by a
Doppler-like asymptotic~(which eventually merges into the Lorentz higher scale asymptotic~\cite{PaperI}). The~$a$ parameter
quantifies the relative importance of Lorentz~(natural plus collisional in this case) over Doppler; the higher the~$a$ value
the more important the Lorentz complete redistribution in the atom's rest frame and therefore the smaller the deviations
from CFR and the onset of the CFR Lorentz asymptotic occurs earlier~(ie., for higher opacity values).

The conditional ODFs for the~$R_{II}$ redistribution are shown in Fig.~\ref{Fig-H-RII}. The most remarkable difference from
the previous cases resides in the low line opacity asymptotic. There is a very abrupt cut-off of the ODF PDF values for
small opacities which scales approximately as~$a$. This case corresponds to excited levels not perturbed by collisions
during their lifetime and coherent reemission in the Atom's Rest Frame~(ARF). This coherent reemission in the ARF does not
give automatically coherence in the Lab Rest Frame~(LRF) due to the Maxwell velocity distribution. However, after several
Doppler widths~(depending on~$a$), the ARF coherency appears in the cut-off of the ODF. As for the previous~$R_{III}$ case,
two regimes can be distinguished, with a transition between them that depends on~$a$. For line opacities approaching the
center-of-line value,~$\Phi_{0}$, the redistribution is similar to the pure Doppler $R_{I}$ case; an initial very steep
(faster than Lorentzian and persisting to smaller line opacities the more off-center is the previous photon absorption)
increase with decreasing opacity value followed by a CFR Doppler redistribution until the small opacities approaching the
cut-off values~(compare the CFR asymptotic with the ODF curves for the~$x^{\prime}=0$, $1$ and $2$ cases). The cut-off of
the ODF PDF for the smaller line opacities is very important since it shows the breakdown of the superdiffusion character
for this~$R_{II}$ model.

From the above results some general conclusions can be extracted:~(i)~the ARF frequency coherence does not immediately give
rise to LRF coherence due to the ensemble Maxwellian velocity distribution averaging and (ii)~the ARF redistribution (or non
redistribution) character will eventually show up as a corresponding LRF asymptotic for sufficiently small line
opacities~(that is to say, for optically thick enough vapors). The CFR in the ARF for the~$R_{III}$ will give rise to a
Lorentz CFR (superdiffusive) asymptotic in infinite media even under partial redistribution while the complete coherence in
the ARF for~$R_{II}$ will reveal itself in the LRF as a breakdown of the superdiffusive charater. The breakdown of the
superdiffusion for thick vapors under~$R_{II}$ conditions is particularly important due to the widespread practice of
reducing the actual complex behavior of the~$R_{II}$ redistribution into a linear superposition of complete frequency
redistribution in the line core plus a complete coherence in the line wings. This practice is essentially motivated by the
mathematical simplification that it allows and it can be traced back to the work of Jefferies and White~\cite{JW}~(and it is
therefore known as the JW-approximation). Although the several implementations of the JW-approximation differentiate the
photons from the core from those from the wings, all ignore the possibility that even the most extreme wing photons can be
redistributed in frequency, thus neglecting the so-called ``difusion'' in frequency. Moreover, the full redistribution
function is skewed towards the line center and this is not taken into account. Figure~\ref{Fig-H-RII} clearly shows another
physical limitation of the JW-approximations which is generally not pointed out. For high opacity samples, the CFR
assumption in core radiation is a severe approximation since CFR implies that radiation well into the wings can be emitted
in a single scattering event, which can not occur (an upper limit exists) due to the breakdown of the Doppler asymptotic.
This contributes to show that the JW-type approximations are only acceptable for small Voigt parameters and for overall
center-of-line opacities roughly smaller than~$100-500$~\cite{Molisch1998,Post1986,Anderson1995}.

The asymptotic analysis of resonance radiation trapping started probably with Holstein seminal work~\cite{Holstein1947} and
received afterwards a relevant contribution from van~Trigt's infinite opacity expansion analysis in the
1970s~\cite{vanTrigt1970s}. Asymptotic analysis is a very useful tool to calculate the large-scale behavior, which, from the
derived scaling laws, can provide physical insight into more complex and realistic conditions. This is particularly true for
partial redistribution in the~LRF. We clearly show that the large scale behavior is of the CFR Lorentz superdiffusive type
for complete redistribution in the~ARF but of the diffusive character if the excited atom is intrinsically coherent~(even if
the small scale resembles the pure Doppler superdiffusive character). So, one can use a generalized diffusive transport
equation~\cite{Metzler2000and2004} for line radiation in this case, even if the effective mean transport coefficients will
hide the details of the small scale partial redistribution effects. The present work is in this respect complementary of the
asymptotic analysis of the integral equation of line radiation transfer carried out by Frisch~\cite{Frisch1980,Frish1985A}.
The work presented here is much simpler and straightforward than Frisch's approach. Furthermore, we clearly identify the CFR
in the line core JW-type approximation for~$R_{II}$ as incompatible with the large scale diffusive behavior.

\begin{figure}
  \includegraphics[width=10cm,keepaspectratio=true]{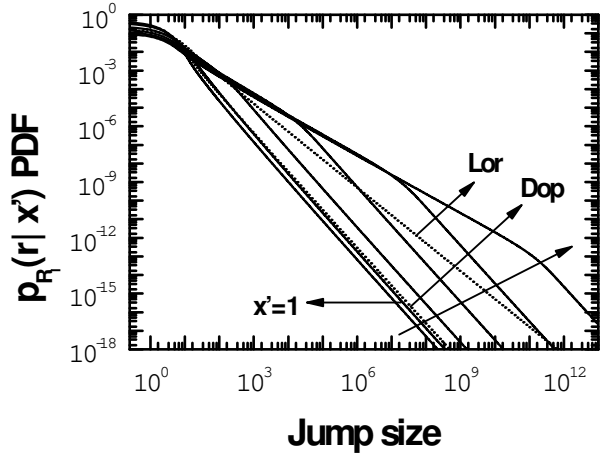}
  \caption{\label{Fig-Jump-RI}
    Conditional one-sided Jump Probability Distribution Function~(PDF) for Partial Redistribution function~$R_{I}$ for several
    previous absorption frequencies~$x^{\prime}$ in the sequence~$x^{\prime}=0$, $1$, $2$, $3$, $4$ and $5$~(arrow direction).
    The Complete Redistribution limiting Doppler and Lorentz cases are also shown.}
\end{figure}

\begin{figure}
  \includegraphics[width=10cm,keepaspectratio=true]{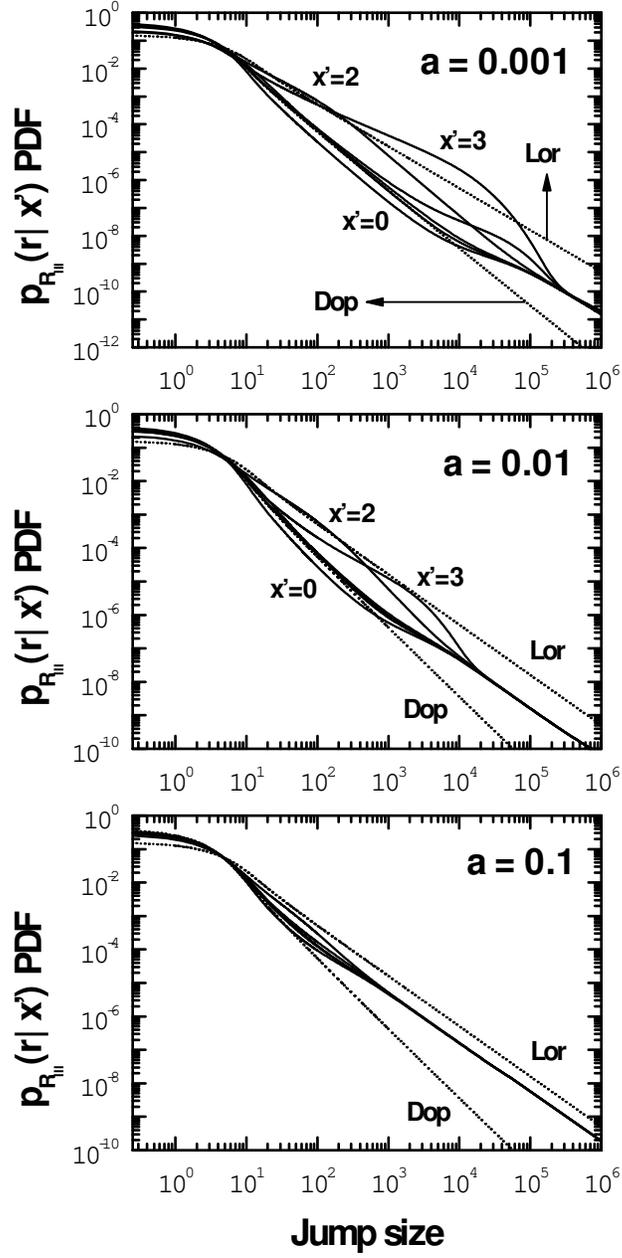}
  \caption{\label{Fig-Jump-RIII}
    Conditional one-sided Jump Probability Distribution Function~(PDF) for Partial Redistribution function~$R_{III}$ for several
    previous absorption frequencies~($x^{\prime}=0$, $1$, $2$, $3$, $4$ and $5$) and Voigt width parameters~$a$. The Complete Redistribution
    limiting Doppler and Lorentz cases are also shown.}
\end{figure}

\begin{figure}
  \includegraphics[width=10cm,keepaspectratio=true]{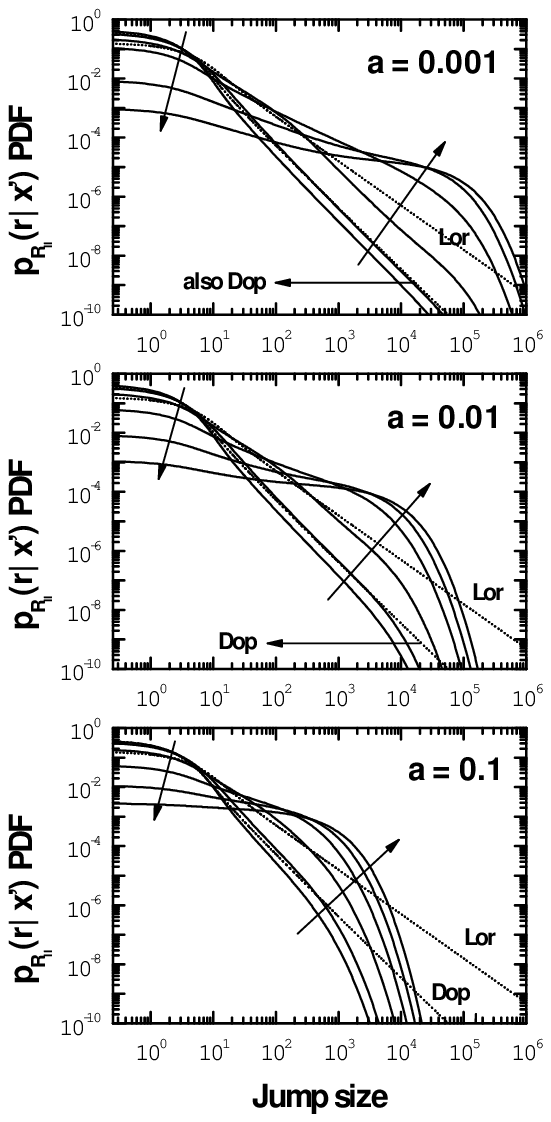}
  \caption{\label{Fig-Jump-RII}
    Conditional one-sided Jump Probability Distribution Function~(PDF) for Partial Redistribution function~$R_{II}$ for several
    previous absorption frequencies~$x^{\prime}$ and Voigt width parameters~$a$. The arrow direction shown results
    in the sequence~$x^{\prime}=0$, $1$, $2$, $3$, $4$ and $5$. The Complete Redistribution
    limiting Doppler and Lorentz cases are also shown.}
\end{figure}

Figures~\ref{Fig-Jump-RI} to~\ref{Fig-Jump-RII} give the jump size PDFs computed numerically from
Eq.~(\ref{eq-PDFprPFR-xline}). These give essentially the same information as the ODF PDFs, although in terms of the overall
opacity length which is directly related to physical distances~(for an additional note on the relation of opacity scales and
physical distances, see ref.~\cite{PaperI}). The superdiffusive behavior of~$R_{I}$ and~$R_{III}$ is evident, while the
breakdown of superdiffusion is manifested by the abrupt cut-off for~$R_{II}$ at higher jump lengths. Smaller details can
also be pointed out. For~$R_{I}$, there is an initial higher than superdiffusive Lorentz jump PDF which changes into the
large scale Doppler asymptotic at high distances~(transition not shown but confirmed for the last two cases
of~$~x^{\prime}=4$ and~$5$ in Fig.~\ref{Fig-Jump-RI}). For the~$R_{III}$ redistribution, all cases converge into the same
Lorentzian CFR. And finally, for the~$R_{II}$ case there are three characteristic jump length ranges; a first in which the
jump PDF increases faster for decreasing opacities than CFR Lorentz, an intermediate with a CFR Doppler character~(and these
two distance ranges roughly correspond to the pure Doppler~$R_{I}$ case) and a third one corresponding to the diffusive
onset due to a step cut-off for higher distances.

The jump size PDFs define the overall opacity scale in which partial frequency redistribution effects are important as well
as the transition from superdiffusion-like into standard diffusion for the~$R_{II}$ case. From Fig.~\ref{Fig-Jump-RIII} one
expects that the onset of CFR Lorentz asymptotic for high opacities should roughly scale with the width~$a$ parameter; PFR
effects should manifest themselves at distances up to~$10^{5}$, $10^{4}$ and $10^{3}$ for~$a=0.001$, $0.01$ and $0.1$,
respectively, when measured in a dimensionless overall opacity scale. On the other hand, in the case of~$R_{II}$
redistribution, Fig.~\ref{Fig-Jump-RII} shows that the standard diffusion should become noticeable only at very high
opacities; roughly~$10^{6}$, $10^{5}$ and $10^{4}$ for~$a=0.001$, $0.01$ and $0.1$, respectively. At smaller scales, and
notably for finite systems with the biggest dimension not exceeding these opacity values, the faster than standard diffusion
behavior is dominant.

\section{DISCUSSION AND CONCLUSIONS}

The line Opacity Distribution Function~(ODF) approach was used to discuss the superdiffusive character of resonance line
radiation transfer under Partial Frequency Redistributions~(PFR) effects, as well as the conditions for its breakdown. These
effects are the result of the ensemble average from the atom's rest frame~(ARF) redistribution function into the lab
frame~(used to describe radiation transport). We discuss PFR effects for three frequency redistribution functions between
absorption and reemission:~$R_{I}$~(pure Doppler broadening), $R_{II}$~(Doppler plus natural) and~$R_{III}$~(combined
Doppler, natural and collisional broadening). The Doppler ARF of~$R_{I}$ will give rise to the Complete Frequency
Redistribution~(CFR) superdiffusive Doppler asymptotic for sufficiently small line opacities, while the Lorentz ARF
intrinsics of~$R_{III}$ give also small line opacity large scale Lorentzian CFR superdiffusive asymptotics. However, the ARF
frequency coherence of~$R_{II}$ gives rise to a different behavior, because the abrupt cut-off in the jump size distribution
for higher distances means that there is a breakdown of the superdiffusive behavior. This is not recognized in both the
astrophysical spectral line formation as well as in the lab scale atomic vapors studies. This breakdown means that the
popular JW-type approximations of PFR under~$R_{II}$ conditions are a severe approximation, since they assume that core
radiation CFR allows a single scattering to photons well into the absorption wings. Under experimental conditions
rendering~$R_{II}$ realistic, a generalized diffusion equation should be valid even for core line radiation. This has a
classical analog found in physical kinetics, where the macroscopic hydrodynamic behavior is derived from a microscopic
description through a Chapman-Enskog expansion~\cite{Chapman1970}.

The~$R_{II}$ and~$R_{III}$ redistribution functions assume either a complete frequency coherence or a complete
redistribution in the~ARF, respectively. These situations should correspond to two limiting cases for which the collisions
during the excited state lifetime are almost nonexistent or very frequent. For resonance lines, the general form of the
redistribution function should be~\cite{OSC1972,Frish1985A,Post1986}

\begin{equation}
  \label{eq-R-general}
  R\left( x^{\prime };x\right) =
  \left( 1-P_{c} \right) \, R_{II} \left( x^{\prime };x\right) +P_{c} \, R_{III}\left( x^{\prime };x\right)
  \mbox{,}
\end{equation}

where the branching ratio~$P_{c} =\frac{\Gamma_{c}}{\Gamma_{c}+\Gamma_{rad}}$ is defined as the probability that an elastic
collision will destroy the correlation between the absorbed and the reemitted frequencies in the~ARF and should go to one in
the limit of high densities and to zero in the low density limit. This equation reflects the competition between complete
redistribution in the~ARF on the one hand and almost complete coherence on the other hand, depending on the elastic
collisional rate. Presumably, below some critical value for~$P_{c}$,~$R_{II}$-type redistribution will dictate a large scale
diffusive behavior while after that critical value, $R_{III}$ will take control of the large scale behavior giving rise to
superdiffusion. The transition condensed in Eq.~(\ref{eq-R-general}) is important since the classical landmark work of Post
in the 1980s for trapping under~PFR with 185~nm Hg radiation~\cite{Post1986} corresponds to a range of~$P_{c}$ values
roughly from 1\% to 30\%.

Figs.~\ref{Fig-H-RI} to~\ref{Fig-Jump-RII} display the \textit{conditional} ODFs and jump size distributions for a given
previous absorption optical frequency. The radiation migration in physical space is not a Markov process whenever~PFR is
meaningful since the whole hierarchy of the process cannot be constructed from the spatial distribution of the excitation in
a given initial time plus the single (spatial) transition probability~(contrary to the~CFR case where the spatial transition
probability is independent of past history). However, the spatial evolution of excitation can be embedded in a Markov
process defined as the excitation migration in both spatial \textit{and} frequency space~\cite{vanKampen1992}. This extended
Markov process will consider explicitly the joint redistribution between absorption and reemission frequencies, information
that otherwise would be contained implicitly in the past values of the absorption frequencies. In principle, it would even
be possible to define two different Markov processes for the all the $R$s redistribution mechanisms, depending on the
coarsening level of the description~\cite{vanKampen1992}. On a fine level description, one should have a complete stochastic
formulation of the radiation trapping problem encompassing both the small scale~(controlled by PFR) as well as the large
scale~(asymptotic) behaviors. But one could also devise a coarser level description retaining only the large scale
asymptotic and blurring the small scale PFR details into a transition probability valid only for large distances. For
the~$R_{I}$ and~$R_{III}$ cases, the large scale transition probability would be given by the corresponding Doppler or
Lorentz CFR superdiffusive asymptotic while for the~$R_{II}$ a simpler diffusion-type behavior should apply.

Finally we discuss the applicability of the Opacity Distribution Function~(ODF) formalism to astrophysical applications. In
astrophysics, we have to deal with a very large number of spectral lines that overlap. Typically, consideration of all the
lines (plus continua) in reaction-hydrodynamic modelling is prohibitive, but this can be circumvented by the use of an ODF.
Detailed studies of this approach have shown that ODFs reproduce satisfactorily both emergent fluxes and the physical
stellar atmospheric structure~\cite{Mihalas1978}. In astrophysical radiative transfer problems, each individual opacity will
be the result of grouping several lines. Recently, Wehrse and coworkers have used a Poisson distributed point process for
the statistical description of the number of lines~\cite{Wehrse2001and2002}. However, they did not attempt a comprehensive
investigation of the dependencies of the ODFs on the detailed line profiles. The present work is a contribution towards the
improvement of this approach by including the effect of the absorption and reemission profiles in a similar way used by us
to study the anomalous diffusion resulting from the redistribution along a single line profile.

\section*{ACKNOWLEDGEMENTS}
%


This work was supported by Funda\c{c}\~{a}o para a Ci\^{e}ncia e Tecnologia~(FCT,~Portugal) within
project~POCI/QUI/58535/2004 and by FCT and Universidade do Minho~(Portugal) within project~REEQ/433/EEI/2005. It also used
computational facilities bought under project~POCTI/CTM/41574/2001, funded by FCT and the European Community Fund~FEDER.
A.R.~Alves-Pereira also acknowledges~FCT funding under the reference~SFRH/BD/4727/2001.







\end{document}